\documentstyle[aps,preprint,amssymb,psfig]{revtex}
\begin{document}
\draft
\preprint{\vbox{\it to appear in Phys. Rev. D \hfill\rm TRI-PP-95-58}}

\title{\bf Chiral symmetry breaking and cooling in lattice QCD}
\author{R.M. Woloshyn and Frank X. Lee}
\address{TRIUMF, 4004 Wesbrook Mall,
Vancouver, British Columbia, Canada V6T 2A3}
\date{\today}
\maketitle

\begin{abstract}
Chiral symmetry breaking is calculated as a function of cooling 
in quenched lattice QCD. A non-zero signal is found for the 
chiral condensate beyond one hundred cooling steps, suggesting 
that there is chiral symmetry breaking associated with instantons.
Quantitatively, the chiral condensate in cooled gauge field 
configurations is small compared to the value without cooling.
\end{abstract}
\vspace{1cm}
\pacs{PACS numbers: 12.38.Gc, 11.15.Ha, 11.15Kc}

Recently Chu {\it et al}~\cite{chuetal} presented evidence from a quenched
SU(3) lattice gauge theory that instantons play a dominant role in
determining hadronic correlation functions. Motivated by this work,
a calculation was done in quenched SU(2) lattice gauge theory exploring
the effect of cooling~\cite{coolsu2}. In Ref.~\cite{coolsu2} it was found
that the spin-dependent potential and the chiral symmetry breaking order
parameter disappear very quickly upon cooling.
In a subsequent nonquenched SU(3) calculation
by Edwards {\it et al}~\cite{edward}, it was found that cooling has a large
effect on the hadron spectrum, tending, for example, to diminish
spin-slittings.
This supports the conclusions of Ref.~\cite{coolsu2}.
However, Edwards {\it et al}~\cite{edward}
did not see the rapid
disappearance of the chiral order parameter found in the quenched
SU(2) calculation. For this reason we have decided to revisit the problem
of chiral symmetry breaking and cooling. A calculation was done in
quenched SU(3) lattice gauge theory to see if the different chiral symmetry
breaking pattern observed in Ref.~\cite{coolsu2} and in
Edwards {\it et al}~\cite{edward}
is due to the use of SU(2) versus SU(3) color or due to the
quenched versus nonquenched nature of the simulation.

In this paper we present results for the chiral order parameter
calculated in a set of quenched SU(3)-color gauge configurations to
which local cooling~\cite{berg,teper1}
 has been applied. A persistent non-vanishing
signal for chiral symmetry breaking was found even for cooling of
a hundred sweeps or more. This differs from what was observed
in the earlier quenched SU(2) simulation~\cite{coolsu2} but is consistent
with the finding of Edwards {\it et al}~\cite{edward}
in their nonquenched SU(3) calculation.
Quantitatively, after cooling, the chiral order parameter is small
(in lattice units) compared to the uncooled value.

The calculations were done using standard methods of Ref.~\cite{coolsu2}
carried over to SU(3).
The fermion action is
\begin{eqnarray}
   S_f & = & \frac{1}{2} \sum_{x,\mu} \eta_\mu(x)
   \left[ \overline{\chi}(x) \, U_\mu(x)
   \,\chi(x+\hat{\mu}) \,\, -\overline{\chi}(x+\hat{\mu})
   U^{\dagger}_\mu(x) \, \chi(x) \right]
   + \sum_x \, m \overline{\chi}(x) \chi(x),
\nonumber \\
   & \equiv & \overline{\chi} \,\, {\cal M}(\{U\}) \, \chi  ,
\label{Sf}
\end{eqnarray}
where $\overline{\chi}, \chi$ are single-component fermion fields,
$\eta_\mu(x)$ is the staggered fermion phase ~\cite{kawo},
$m$ is the mass (in units of the inverse of the lattice spacing $a$)
and the $U's$ are gauge field links.
Antiperiodic boundary conditions were used for the staggered fermion
fields in all directions.

The chiral symmetry order parameter is calculated from the
inverse of the fermion matrix ${\cal M}$ of Eq.~(\ref{Sf})
\begin{eqnarray}
   \langle \overline{\chi} \chi \rangle
   = \frac{1}{V} \, \langle \mbox{Tr} \, {\cal M}^{-1}(\{U\}) \rangle ,
\end{eqnarray}
where V is the lattice volume and the angle brackets
indicate a gauge field configuration average. The random source
method ~\cite{scalet} was used to calculate
$\mbox{Tr} {\cal M}^{-1}(\{U\})$. Twelve Gaussian random sources
were used for each gauge field configuration.

Calculations were done on a $10^4$ lattice at $\beta = 5.7$ and on a
$12^4$ at $\beta = 5.9$. The SU(3) pseudo-heatbath method was used for
updating the quenched gauge field configurations. After 4000 sweeps
to thermalize, configurations were analyzed every 300 sweeps. A total
of 30 configurations at $\beta = 5.7$ and 20 configurations at
$\beta = 5.9$ were included in the analysis. Each configuration
was cooled using a local cooling algorithm~\cite{berg,teper1}
and $\langle \overline{\chi}
\chi \rangle$ was calculated along the way.

\begin{figure}
\centerline{\psfig{file=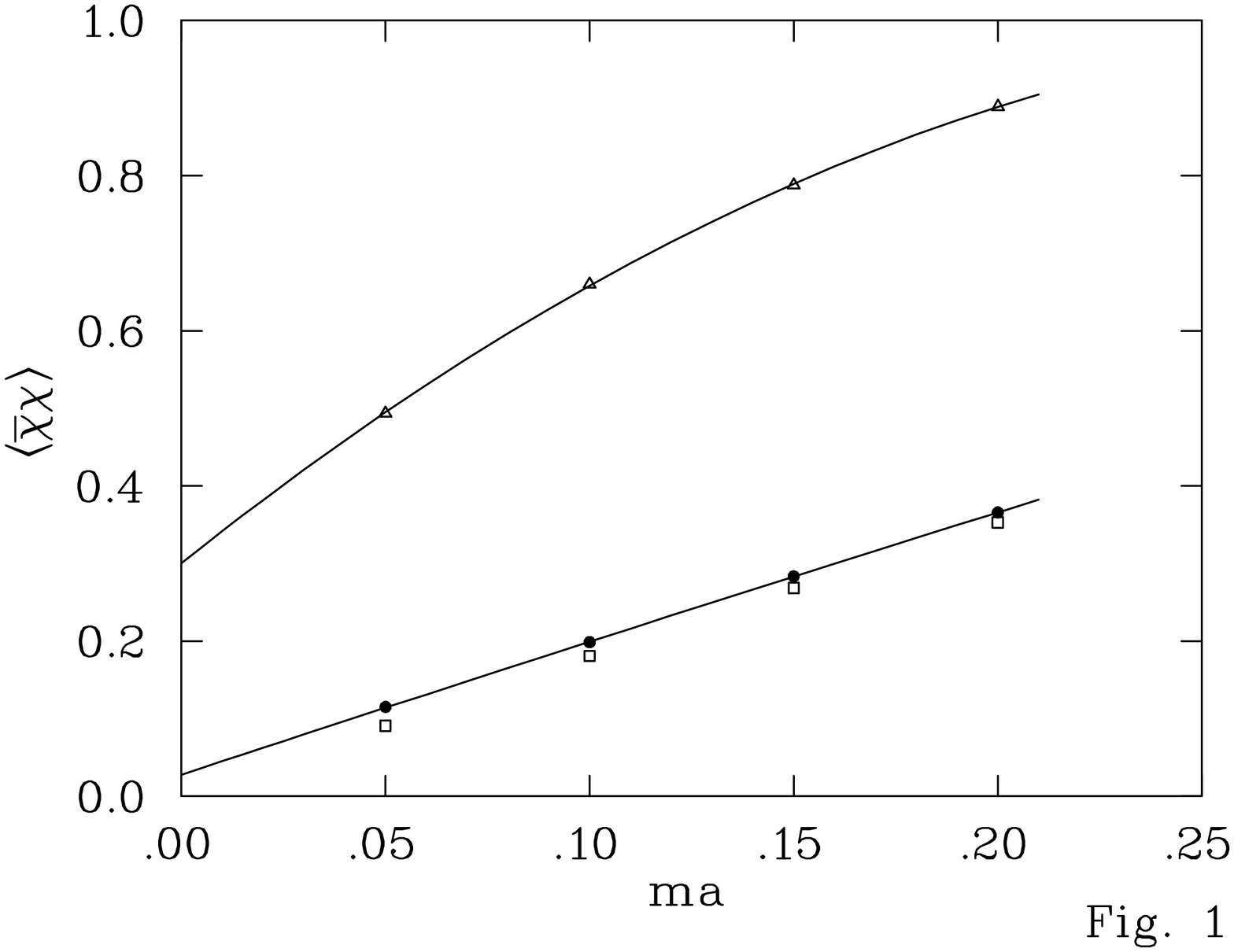,width=14cm,angle=90}}
\vspace{1cm}
\caption[dum1]{Plot of the chiral order parameter (in lattice units)
as a function of quark mass at $\beta = 5.7$ 
for no cooling ($\triangle$) and 20 cooling steps ($\bullet$). 
The lattice free-field values are also shown ($\Box$).
The lines are the extrapolations to zero quark mass using 
Eq.~(\ref{trap}).}
\label{b57}
\end{figure}

\begin{figure}
\centerline{\psfig{file=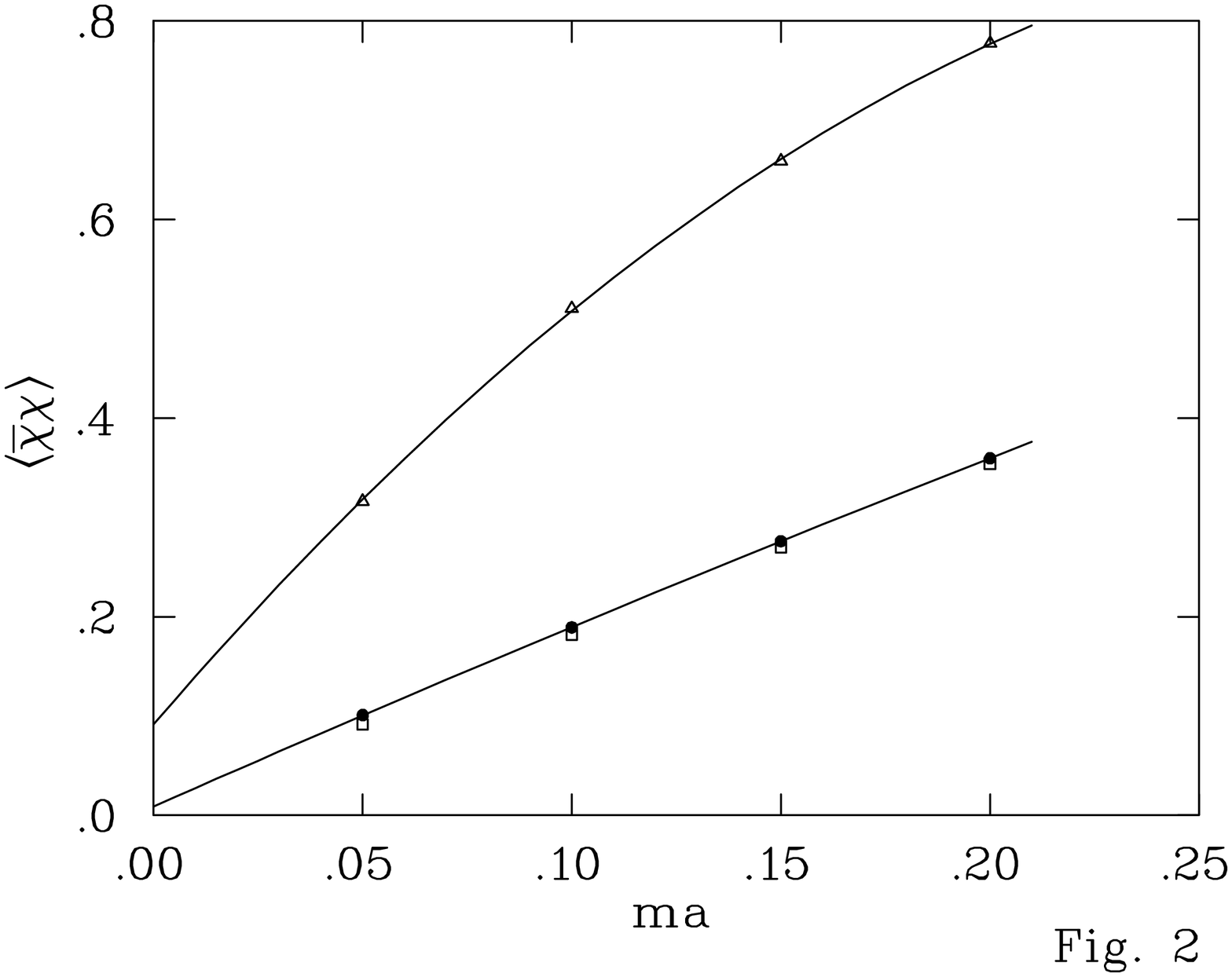,width=14cm,angle=90}}
\vspace{1cm}
\caption[dum2]{Same as in Fig.~\ref{b57}, except for $\beta = 5.9$.}
\label{b59}
\end{figure}

The results are tabulated in Table~\ref{extra}. 
The results for $\langle \overline{\chi}
\chi \rangle$ as a function of quark mass
at no cooling and after
20 cooling sweeps are shown in Fig.~\ref{b57} and Fig.~\ref{b59} 
for $\beta = 5.7$ and $\beta = 5.9$ respectively. 
The value of $\langle \overline{\chi} \chi \rangle$
obtained in a free field theory is also plotted in the figures.

\begin{figure}
\centerline{\psfig{file=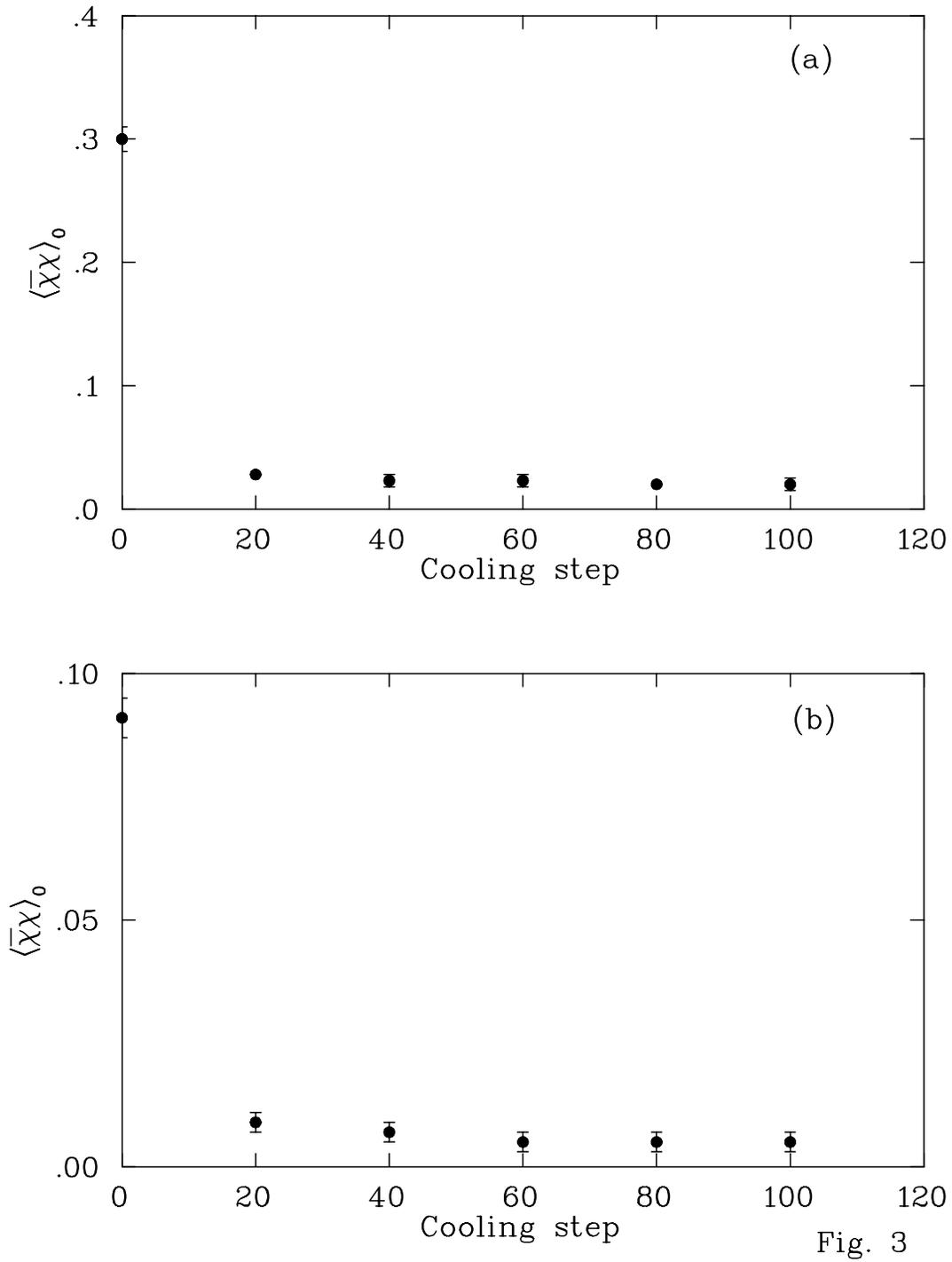,width=14cm}}
\vspace{1cm}
\caption{The chiral order parameter at zero quark mass
$\langle \overline{\chi}\chi \rangle_0$ versus number
of cooling sweeps at (a) $\beta = 5.7$, and (b) $\beta = 5.9$.}
\label{cool}
\end{figure}

Following Bowler {\it et al}~\cite{bowler}, the values at zero
quark mass (extrapolated values)
were obtained by fitting to the function
\begin{equation}
   \langle \overline{\chi}\chi \rangle(m)
   = \langle \overline{\chi}\chi \rangle_0
   + \langle \overline{\chi}\chi \rangle_1 \, m
   + \langle \overline{\chi}\chi \rangle_2 \, m^2 .
\label{trap}
\end{equation}
The resulting chiral condensate values 
$\langle \overline{\chi}\chi \rangle_0$
are plotted in Fig.~\ref{cool} versus the number of cooling sweeps.
A clear nonzero signal for chiral symmetry breaking is obtained 
confirming the conclusion of Edwards {\it et al}~\cite{edward}.
This is in clear contrast
to the results of Ref.~\cite{coolsu2} (see Figs. 9-11) where the chiral
order parameter was found to disappear very quickly with cooling. It is
fair to say, however, that the value of the chiral order parameter 
found in our calculation,
although nonzero after cooling, is small compared (in lattice units) to
the uncooled value. 

\begin{figure}
\centerline{\psfig{file=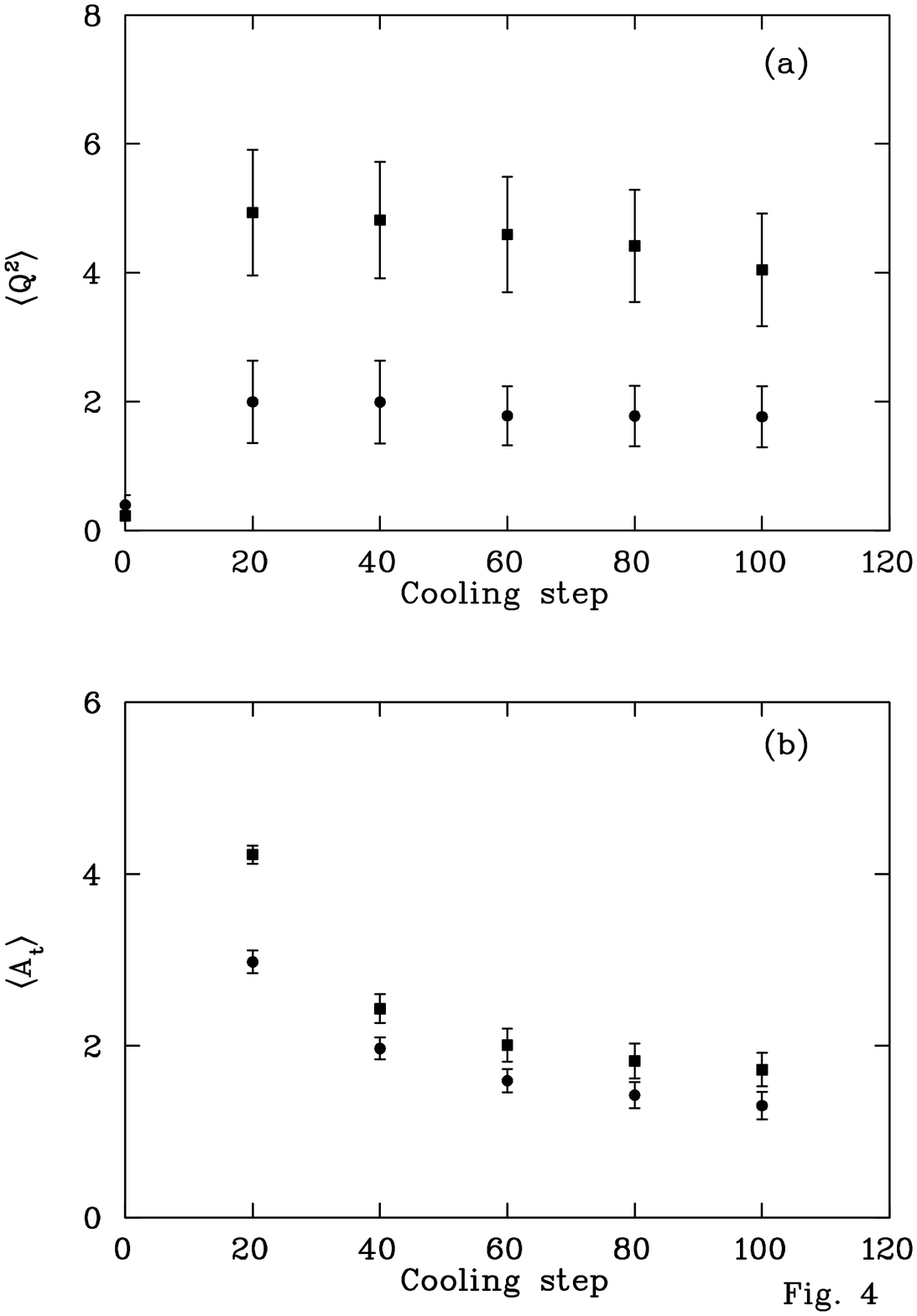,width=14cm}}
\vspace{1cm}
\caption{Expectation value of topological charge squared (a) 
and topological activity (b) as a function of cooling at 
$\beta = 5.7$ ($\blacksquare$) and $\beta = 5.9$ ($\bullet$).}
\label{topo}
\end{figure}

The behavior of topological properties under cooling is well
known~\cite{poli}. Here we give a brief description to put our 
calculation into context. The simple transcription of 
$F_{\mu\nu}(x)\tilde{F}_{\mu\nu}(x)$ to the lattice~\cite{di} is 
used for the topological charge density
\begin{equation}
q_t(x)={1\over 32\pi^2}\sum_{\mu\nu\rho\sigma}\epsilon_{\mu\nu\rho\sigma}
\mbox{Tr}\{ U_{\mu\nu}(x) U_{\rho\sigma}(x)\},
\end{equation}
where $U_{\mu\nu}$, $U_{\rho\sigma}$ are plaquettes of gauge field links.
Fig.~\ref{topo} shows the expectation values of topological charge 
squared $Q^2$ ($Q=\sum_x q_t(x)$) and a measure of the topological 
activity, $A_t=\sum_x|q_t(x)|$.
Note that without cooling the topological activity $A_t$ is very large
due to short distance fluctuations in $q_t(x)$. The numbers are too
large to be plotted in Fig.~\ref{topo}(b) and are $45.02\pm0.05$ at 
$\beta=5.7$ ($10^4$ lattice) and $87.6\pm0.1$ at $\beta=5.9$ 
($12^4$ lattice).
In the first ten or so cooling steps, short distance fluctuations 
disappear very quickly, the topological activity drops rapidly and 
$\langle Q^2 \rangle$ takes a value which changes only slowly with 
more cooling. In the region up to about $60$ cooling steps 
$\langle A_t  \rangle^2$ is larger than $\langle Q^2 \rangle$,
indicating that one does not yet have gauge field configurations 
consisting of single isolated instantons. 
Rather, the vacuum in this region of cooling is sometimes characterized 
in terms of ``overlapping'' or ``interacting'' 
instantons~\cite{chuetal,poli}.  Beyond about $60$ cooling steps 
$\langle A_t \rangle^2 < \langle Q^2 \rangle$
as one would expect for a sample of instanton configurations with 
differing topological charge. The rate at which the topological 
activity changes becomes very small as instantons, which are not 
absolutely stable on the lattice, slowly shrink and fade away. 
By $100$ cooling steps one expects to be in the domain of a vacuum 
dominated by isolated instantons~\cite{chuetal,coolsu2,poli}.
Our finding that the value of the chiral order parameter is 
non-zero and essentially constant beyond $60$ cooling steps, 
we would say, gives an indication that, in an  
SU(3) gauge theory, instantons produce chiral symmetry breaking,
although it would not suggest that instantons dominate 
chiral symmetry breaking.

At present we have no explanation for the different behavior observed in the
SU(2) and SU(3) calculations. One possibility is that the SU(2) calculation,
which was done at $\beta = 2.4$, is just deeper in the weak coupling region
than the SU(3) calculations done here at $\beta = 5.7$ and $5.9$. Indeed,
comparing the SU(3) calculations at the two $\beta$ values may suggest a
diminished persistence of chiral symmetry breaking at $\beta = 5.9$
but this conclusion can not be made definitively due to the smallness of the
chiral condensate (and concomitant large uncertainty)
 at the larger $\beta$ value. It may also be that there is
still some more fundamental difference between the SU(2) and SU(3) theories
which has yet to be found.

It is a pleasure to thank H.D. Trottier
for helpful discussions.
This work was supported in part by the Natural Sciences and
Engineering Research Council of Canada.

\begin{table}
\caption{Chiral order parameter values calculated for different
values of the quark mass along with the value 
extrapolated to zero mass.}
\label{extra}
\begin{tabular}{ccccccc}
 && \multicolumn{4}{c}{ma} & \\ \cline{3-6} 
$\beta$ & Cooling step 
& 0.2    & 0.15      &  0.1      & 0.05     & 
Extrapolated value \\ \hline
5.7 & 0  & 0.889(2)  & 0.788(2)  & 0.660(2)  & 0.494(3) & 0.30(1) \\
    & 20 & 0.3659(3) & 0.2835(5) & 0.1986(7) & 0.115(2) & 0.028(4) \\
    & 40 & 0.3613(5) & 0.2783(5) & 0.1930(8) & 0.110(2) & 0.023(5) \\
    & 60 & 0.3600(4) & 0.2768(6) & 0.1913(9) & 0.109(2) & 0.020(4) \\
    & 80 & 0.3593(5) & 0.2761(6) & 0.1907(9) & 0.107(2) & 0.020(4) \\
    & 100& 0.3592(5) & 0.2757(6) & 0.1899(9) & 0.107(2) & 0.020(5) \\
[0.5cm]
5.9 & 0  & 0.7775(7) & 0.6589(8) & 0.5104(9) & 0.316(1) & 0.091(4) \\
    & 20 & 0.3595(2) & 0.2760(2) & 0.1892(3) & 0.1009(5)& 0.009(2) \\
    & 40 & 0.3583(2) & 0.2745(2) & 0.1874(3) & 0.099(6) & 0.007(2) \\
    & 60 & 0.3576(2) & 0.2738(2) & 0.1866(3) & 0.098(6) & 0.005(2) \\
    & 80 & 0.3576(2) & 0.2736(2) & 0.1862(3) & 0.098(6) & 0.005(2) \\
    & 100& 0.3573(2) & 0.2732(2) & 0.1859(3) & 0.097(6) & 0.005(2) \\
\end{tabular}
\end{table}

\end{document}